\documentclass[12pt,showpacs]{revtex4}

\def\comment#1{}
\usepackage{amsmath}
\usepackage{amssymb}
\usepackage{graphicx}
\usepackage{epsfig}
\def\beq{\begin{equation}}
\def\eeq{\end{equation}}
\def\bea{\begin{eqnarray}}
\def\eea{\end{eqnarray}}
\def\no{\nonumber}

\usepackage{color}
\def\red#1{\textcolor{red}{#1}}
\def\blu#1{\textcolor{blue}{#1}}

\def\mn#1{\marginpar[\tiny{\red{#1}}]{\tiny{\blu{#1}}}}

\begin{document}

\title{Electron and positron pair production of compact stars}

\author{Wen-Biao Han,$^{a,b}$ Remo Ruffini,$^{a}$ She-Sheng Xue$^{a}$}
\affiliation{$^{a}$
ICRANet, P.zza della Repubblica 10, I--65122 Pescara, Physics Department 
and 
ICRA, University of Rome {\it La Sapienza}, P.le Aldo Moro 5, I--00185 Rome, Italy.\\
$^{b}$ Shanghai Astronomical Observatory, 80 Nandan Road, Shanghai, 200030, China.}

\date{Received version \today}

\begin{abstract}
Neutral stellar core at or over nuclear densities is described by a positive charged baryon core and negative charged electron fluid since they possess different masses and interactions. 
Based on a simplified model of a gravitationally collapsing or 
pulsating baryon core, we approximately integrate the Einstein-Maxwell equations and 
the equations for the number and
energy-momentum conservation of complete degenerate electron fluid. We show possible electric processes that 
lead to the production of electron-positron pairs  
in the boundary of a baryon core and calculate the number and energy of electron-positron pairs. This can be relevant for understanding the
energetic sources of supernovae and gamma-ray bursts.
\end{abstract}
\maketitle

\section{\bf Introduction.}\label{int_s}

In the gravitational collapse or pulsation of neutral stellar cores at densities comparable to the nuclear density, complex dynamical processes are expected to take place. These involve both macroscopic processes such as gravitational and hydrodynamical processes, as well as microscopic processes due to the strong and electroweak interactions. The time and length scales of macroscopic processes are much larger than those of the microscopic processes. Despite the existence of only a few exact solutions of Einstein's equations for simplified cases, 
macroscopic processes can be studied rather well by numerical algorithms. 
In both analytical solutions and numerical simulations it is rather difficult to simultaneously analyze both macroscopic and microscopic processes characterized by such different time and length scales.
In these approaches, 
microscopic processes are approximately treated as local and instantaneous
processes that are effectively represented by a model-dependent parametrized equation of state (EOS). We call this {\it approximate locality}.
\comment{ 
}
 
Applying {\it approximate locality} to electric processes, as required by the charge conservation, one is led to {\it local neutrality}\,: positive and negative charge densities are exactly equal over all space and time. As a consequence, all electric fields and processes are eliminated. An internal electric field (charge separation) {\it must} be developed \cite{ob_1975} in a totally neutral system of proton and electron fluids in a gravitational field. If the electric field (process) is weak (slow) enough, {\it approximate locality} is applicable. However, this
should be seriously questioned when the electric field (process) is strong (rapid). For example, neutral stellar cores reach the nuclear density where positive charged baryons interact via the strong interaction while electrons do not, in addition to their widely different masses. As a result, their pressure, number, and energy density are described by different EOS, and a strong electric field
(charge separation) on the baryon core surface is realized \cite{Usov1} in an electrostatic equilibrium state. 
\comment{These have been clarified in static equilibrium configurations, as will be briefly discussed below.
Furthermore, triggered by
} 

Furthermore, either gravitationally collapsing or pulsating of the baryon core leads to the dynamical evolution of electrons. As a consequence, the strong electric field dynamically evolves in space and time, and some electromagnetic processes can result if their reaction rates are rapid enough, for example, the electron-positron pair-production process of Sauter-Heisenberg-Euler-Schwinger (see the review \cite{phreport}) for electric fields $E\gtrsim E_c\equiv m_e^2c^3/(e\hbar)$. 
\comment{
}
If this indeed occurs, gravitational and pulsating energies of neutral stellar cores are converted into the observable energy of electron-positron pairs via the space and time evolution of electric fields. In this article, we present our studies of this possibility
(the natural units $\hbar=c=1$ are adopted, unless otherwise specified).

\section{\bf Basic equations for dynamical evolution.}\label{basic}

We attempt to study possible electric processes in the dynamical perturbations of neutral stellar cores. These dynamical perturbations can be caused by either the gravitational collapse or pulsation of neutral stellar cores.
The basic equations are the Einstein-Maxwell equations and those governing the particle number and
energy-momentum conservation 
\begin{eqnarray}
\quad\quad\,\,\,\,\,\,\,\,  (\bar n_{e,B}U^\nu_{e,B})_{;\nu}&=& 0,\no\\
G_{\mu\nu} &=&  - 8 \pi G (T_{\mu\nu}+T^{\rm em}_{\mu\nu}),\no\\ (T^{\nu}_{\,\,\,\mu})_{;\nu} &=& -F_{\mu\nu}J^{\nu},\nonumber\\
F^{\mu\nu}_{\,\,\,\,\,\,\,;\nu} &= &4\pi J^{\mu},
\label{coeqns1}
\end{eqnarray}
in which the Einstein tensor $G_{\mu\nu}$,  the electromagnetic field $F^{\mu\nu}$ (satisfying $F_{[\alpha \beta, \gamma]}=0$) and its energy-momentum tensor $T^{\rm em}_{\mu\nu}$ appear; $U^\nu_{e,B}$ and $\bar n_{e,B}$ are, respectively, the four velocities and proper number-densities of the electrons and baryons.   
The electric current density is
\begin{eqnarray}
J^\mu= e\bar n_pU^\mu_{B}-e\bar n_eU^\mu_{e},
\label{current}
\end{eqnarray}
where $\bar n_p$ is the proper number-density of the positively charged baryons. 
The energy-momentum tensor $T^{\mu\nu}=T_e^{\mu\nu}+T_B^{\mu\nu}$ is taken to be that 
of two simple perfect fluids representing the electrons and the baryons, 
each of the form
\begin{eqnarray}
T^{\mu\nu}_{e,B} &=& \bar p_{e,B}g^{\mu\nu}+(\bar p_{e,B}+\bar\rho_{e,B})U^\mu_{e,B} U^\nu_{e,B},
\label{etensor}
\end{eqnarray}
where $\bar\rho_{e,B}(r,t)$ and $\bar p_{e,B}(r,t)$ are the respective proper energy
densities and pressures. 

In this article, baryons indicate hadrons, or their constituents (quarks) that carry baryon numbers. Electrons indicate all negatively charged leptons. Baryon fluid and electron fluid are separately described for the reason that in addition to baryons being much more massive than electrons, the EOS of baryons $\bar p_B=\bar p_B(\bar\rho_B)$ is very different from the electron one $\bar p_e=\bar p_e(\bar\rho_e)$ due to the strong interaction. 
Therefore, in the dynamical perturbations of neutral stellar cores, one should not expect that the space-time evolution of number density, energy density, four velocity, and pressure of baryon fluid be identical to the space-time evolution of counterparts of electron fluid. The difference of space-time evolutions of two fluids results in the electric current (\ref{current}) and field $F^{\mu\nu}$, possibly leading to some electric processes. 
In a simplified model for the dynamical perturbations of neutral stellar cores, we approximately study possible electric processes
\comment{ 
} 
by assuming that the equilibrium configurations of neutral stellar cores are initial configurations.

\section{\bf Equilibrium configurations.}

In Refs.~\cite{ob_1975}, the equilibrium configurations of neutral stellar cores, whose densities are smaller than nuclear density $n_{\rm nucl}$, are studied on the basis of hydrostatic dynamics of baryon and electron fluids in the presence of long-ranged gravitational and Coulomb forces. In these equilibrium configurations, very weak electric fields $E\ll E_c$ are present, resulted from the balance between attractive gravitational force and repulsive Coulomb force. This electric field is too weak to make important electric processes, for example, electron-positron pair productions.  We are interested in the case where strong electric fields are present. This leads us to consider strong electric fields in the surface layer of baryon cores of compact stars (quark or neutron stars) at or over the nuclear density. 
In this case, we assume that baryons form
a rigid core of radius $R_c$ and density
\begin{eqnarray}
\frac{\bar n_{B,p}(r)}{\bar n_{B,p}}=\left[\exp\frac{r\!-\!R_c}{\zeta}+1\right]^{-1},\, \bar n_{B,p}\approx \frac{N_{B,p}}{(4\pi R_c^3/3)},
\label{baryon}
\end{eqnarray}
where $\bar n_p/\bar n_B\approx N_p/N_B<1$, $N_B(N_p)$ is the
number of total (charged) baryons and $\bar n_{B,p}\gtrsim n_{\rm nucl}\approx 1.4\times 10^{38}{\rm cm}^{-3} $.
\comment{ 
}
The baryon core has a sharp boundary ($r\sim R_c$) of the width $\zeta\sim m^{-1}_\pi$ due to the strong interaction. The line element is \cite{B_1971}
\begin{eqnarray}
&&ds^2 =-g_{tt}dt^2+g_{rr}dr^2+r^2d\theta^2 +r^2\sin^2\theta
d\phi^2 ~, \label{sw}\\
&&g^{-1}_{rr}(r) = 1 - 2 G M(r)/r + G Q^2(r)/r^2\, ,\nonumber
\end{eqnarray}
where mass $M(r)$, charge $Q(r)$ and radial electric field $E(r)=Q(r)/r^2$.
\comment{ 
We further assume that 
the baryon core is completely rigid, infinite stiff surface?? $(\xi\equiv r-R_c\sim 0)$
We start with a massive baryon core at the nuclear density $n_{\rm nucl}$, as discussed for compact stars in \cite{Usov1}. 
Due to the strong interaction at the range of the pion Compton length $\lambda_\pi$, baryon cores have a sharp boundary, as given for example by the soliton-like solution \cite{tdLee1}. 
The distribution (\ref{baryon}) is established by the balance between gravity and nucleon pressure; its boundary sharpness is determined and adjusted by the strong interaction at the rate $\tau_{\rm stro}^{-1}\sim m_\pi$, which should not be sensitive to adiabatically collapsing processes.
}

Electrons form a complete degenerate fluid and their density $n^{\rm eq}_e(r)$ 
obeys the following Poisson equation and equilibrium condition \cite{rrx2011}:
\begin{eqnarray}
&&\frac{d^2V_{\rm eq}}{dr^2}+ \left[\frac{2}{r}-\frac{1}{2}\frac{d}{dr}\ln(g_{tt}g_{rr})\right]\frac{dV_{\rm eq}}{dr}\nonumber\\
&&\qquad
= - 4\pi e g_{rr}(\bar n_pU^t_p-n^{\rm eq}_eU^t_e),\label{potential0}\\
&&E^F_e = g^{1/2}_{tt}\sqrt{|P_e^{F}|^2+m_e^2}-m_e-eV_{\rm eq}={\rm const.}\, ,\nonumber
\end{eqnarray}
where  $U_p^t=U_e^t=(1,0,0,0)$, $E^F_e$, and $P_e^{F}=(3\pi^2 n^{\rm eq}_e)^{1/3}$ are the Fermi energy and momentum, $V_{\rm eq}(r)$ and $E_{\rm eq}
=-(g_{rr})^{-1/2}\partial V_{\rm eq}(r)/\partial r$ are the static electric potential and field.
\comment{ 
}
In the ultrarelativistic case $|P_e^{F}|\gg m_e$, we numerically integrate Eq.~(\ref{potential0}) with boundary conditions:
\begin{eqnarray}
n^{\rm eq}_e(r)|_{r\ll R_c} &=& n_B\no\\
n^{\rm eq}_e(r)|_{r\gg R_c} &=& \frac {d n^{\rm eq}_e(r)}{dr}\Big|_{r\gg R_c}=\frac {d n^{\rm eq}_e(r)}{dr}\Big|_{r\ll R_c}=0.
\label{b_con}
\end{eqnarray}
As a result, we obtain on the baryon core boundary $r\approx R_c$,
the nontrivial charge-separation $(n_p-n_e^{\rm eq})/n_B$ and overcritical electric field $E_{\rm eq}/E_c>0$ in a thin layer
of a few electron Compton length $\lambda_e$
[the curves ($t=0$) in Fig.~\ref{E3t}]. This is due to the sharpness boundary ($\zeta\sim m_\pi^{-1}$) of the baryon core (\ref{baryon}) at the nuclear density, as discussed for compact stars \cite{Usov1}. Note that all electronic energy-levels \cite{Hagen}
\begin{eqnarray}
{\mathcal E}_{\rm occupied}=e\int  g_{rr}^{1/2}dr E_{\rm eq}(r)
\label{occupy}
\end{eqnarray}
are fully occupied and  pair-production is not permitted due to Pauli blocking, although electric fields in the surface layer are over critical. We want to understand the space and time evolution of the electric field in this thin layer and its consequence in the dynamical perturbations of baryon cores, which can be caused by either the gravitational collapse or pulsation of baryon cores.

\section{\bf Modeling dynamical perturbations of baryon cores.}\label{model}
 
It is rather difficult to solve the dynamical system (\ref{coeqns1}-\ref{etensor}) with the EOS $\bar p_B=\bar p_B(\bar\rho_B)$ and $\bar p_e=\bar p_e(\bar\rho_e)$ for the gravitational collapse or pulsation of baryon core and electron fluid,
and to examine possible electromagnetic processes. The main difficulty comes from the fact that the time and length scales of gravitational and electromagnetic processes differ by many orders of magnitude. In order to gain some physical insight into the problem, we are bound to split the problem into three parts: (i) first, we adopt a simplified model to describe the dynamical perturbations of baryon cores; (ii) second, we examine how electron fluid responds to this dynamical perturbation of baryon cores; (iii) third, we check whether the resulted strong electric fields can lead to very rapid electromagnetic processes, for example, electron-positron pair production.   

As for the first part, we adopt the following simplified model. Suppose that at the time $t=0$ the baryon core is in the equilibrium configuration (\ref{baryon}) with the radius $R_c$ and
starts dynamical perturbations with an inward velocity $\dot R_c(t)$ or pulsation frequency $\omega_{\rm pulsa}\simeq \dot R_c/R_c$. The rate of dynamical perturbations of baryon cores is defined as $\tau^{-1}_{\rm coll}=\dot R_c/R_c \lesssim c/R_c$. We further assume that in these dynamical perturbations, baryon cores are rigid, based on the argument that as the baryon core density $\bar n_{B,p}$ (\ref{baryon}) increases, the EOS of baryons $\bar p_B=\bar p_B(\bar \rho_B)$ due to the strong interaction is such that the baryon core profile (\ref{baryon}) and boundary width $\zeta\sim m^{-1}_\pi$ are maintained in the nuclear relaxation rate $\tau_{\rm stro}^{-1}\sim m_\pi$, which is much larger than $\tau^{-1}_{\rm coll}$. Thus, due to these properties of strong interaction, the dynamical perturbation of the baryon core induces an inward charged baryon current-density 
\begin{eqnarray}
J_B^r(R_c)=e\bar n_p(R_c)U^r_B(R_c),
\label{cur0}
\end{eqnarray}
on the sharp boundary of baryon core density (\ref{baryon}) at $R_c$, where the baryon density $\bar n_{B,p}(R_c)=0.5\bar n_{B,p}$ and the four-velocity $U^r_B(R_c)\not=0$. We have not yet been able, from the first principle of strong interaction theory, to derive this boundary property (\ref{cur0}) of baryon cores undergoing dynamical perturbations, which essentially are assumptions in the present article, and the boundary density $\bar n_{B,p}(R_c)$ and the boundary four-velocity $U^r_B(R_c)$ are two parameters depending on dynamical perturbations. This is in the same situation that so far one has not yet been able, from the first principle of strong interaction theory, to derive the sharp boundary profile (\ref{baryon}) of baryon core densities of static compact stars \cite{Usov1}. However, we have to point out that the boundary properties (\ref{baryon}) and (\ref{cur0})
of the baryon core undergoing dynamical perturbations are rather technical assumptions for the following numerical calculations of dynamical evolution of electron fluid and electric processes in the Compton time and length scales. These assumptions could be abandoned if we were able to simultaneously make numerical integration of  differential equations for both dynamical perturbations of baryon cores at macroscopic length scale and strong and electric processes at microscopic length scale.

\section{Dynamical evolution of electron fluid}\label{s-field}

In this section, we attempt to examine how the electron fluid around the boundary layer of the baryon core responds to the dynamical perturbations of the baryon core described by the boundary properties (\ref{baryon}) and (\ref{cur0}). Given these boundary properties at different values of baryon core radii $R_c$, we describe electrons and electric fields around the boundary layer of baryon core by Maxwell's equations, the electron number and energy-momentum conservation laws (\ref{coeqns1}) in the external metric field (\ref{sw}). In addition, we assume that the electron fluid is completely degenerate, and its EOS is given by 
\begin{eqnarray}
\bar\rho_e(t,r) &=& 2
\int_0^{P^F_e} p^0 d^3{\bf p}/(2 \pi)^3,\no\\
\bar p_e(t,r)&=& \frac{1}{3} \frac{2}{(2 \pi)^3} \int_0^{P^F_e}
\frac{{\bf p}^2}{p^0} d^3{\bf p}\ ,\label{pt}
\end{eqnarray}
where the single-particle spectrum is $p^0=({\bf p}^2+m_e^2)^{1/2}$ and the Fermi momentum is $P^F_e=(3\pi^2 \bar n_e)^{1/3}$. In the present article, for the sake of simplicity, we set the temperature of electron fluid to be zero and neglect all temperature effects, which may be important and will be studied in future.   
\comment{
The assumption (ii) is based on the fact that the rates for electromagnetic processes are much faster than the rates for gravitational and other hydrodynamic processes ($\tau^{-1}_{\rm coll}=\dot R_c/R_c \lesssim c/R_c$), so that the latter can be considered as adiabatic processes with respect to the former. This will be self-consistently verified.
}

The electron fluid has four velocity $U^\mu_{e}=(U^t, U^r)_{e}$, radial velocity $v_{e}\equiv ( U^r/U^t)_{e}$, $U^t_e=g_{tt}^{-1/2}\gamma_e$ and Lorentz factor $\gamma_{e}\equiv (1+U_rU^r)^{1/2}_{e}=[1+(g_{rr}/g_{tt})v_{e}^2]^{-1/2}$. In the rest frame at a given radius $r$ it has the number density
$n_{e}=\bar{n}_{e}\gamma_{e}$, energy density $\epsilon_{e}=(\bar \rho_{e}+\bar p_{e} v^2_{e})\gamma^2_{e}$, momentum density $
P_{e}=(\bar\rho_{e} + \bar p_{e})\gamma^2_{e} v_{e}$, and $v_{e}=P_{e}/(\epsilon_{e}+\bar p_{e})$.
\comment{ 
} 
In the rest frame, the number and energy-momentum conservation laws for the electron fluid, and Maxwell's equations are given by
\begin{align}
 &\left(n_{e}g_{tt}^{-1/2}
\right)_{,t}+\left(n_{e}v_{e}g_{tt}^{-1/2}\right)_{,r}=0,\label{nconserve}\\
&(\epsilon_{e})_{,t}+\left(P_{e}\right)_{,r}+\frac{1}{2g_{tt}}\left[\frac{\partial
g_{rr}}{
\partial t}P_{e}v_{e}-\frac{\partial
g_{tt}}{
\partial t}(\epsilon_{e}+\bar p_{e})\right] \nonumber \\
 &\qquad 
 =- e n_{e} v_{e}E g^{-1/2}_{tt},\label{econserve}\\
  &\left(P_{e}\frac{g_{rr}}{g_{tt}}\right)_{,t}+\left(\bar p_{e}+P_{e} v_{e}
\frac{g_{rr}}{g_{tt}}\right)_{,r} \nonumber \\
&\qquad
+\frac{\epsilon_{e}+\bar p_{e}}{2g_{tt}}\left(\frac{\partial
g_{tt}}{\partial r}-\frac{\partial g_{rr}}{\partial
r}v^2_{e}\right)=- e n_{e} E g^{-1/2}_{tt},\label{pconserve} \\
&(E)_{,t}=-4\pi e(n_p
v_p-n_{e} v_e)g^{-1/2}_{tt},\label{maxwell}
\end{align}
\comment{
}
where $(\cdot\cdot\cdot)_{,x}\equiv (-g)^{-1/2}\partial (-g)^{1/2}(\cdot\cdot\cdot)/\partial x$, and in the line (\ref{maxwell}),
the boundary velocity $v_{p}$ of the baryon core comes from the baryon current-density (\ref{cur0}). We have the boundary four-velocity $U^r_B$ of the baryon core,
\begin{eqnarray}
v_{p} = v_B\equiv ( U^r/U^t)_{B},\quad U^t_B=g_{tt}^{-1/2}\gamma_B,
\label{v_bar}
\end{eqnarray}
and the Lorentz factor 
\begin{eqnarray}
\gamma_{B}\equiv (1+U_rU^r)^{1/2}_{B}=[1+(g_{rr}/g_{tt})v_{p}^2]^{-1/2},
\label{gama_bar}
\end{eqnarray}
at the baryon core boundary $R_c$.   

In the static case for $v_p=v_e=0$, Eqs.~(\ref{pt}-\ref{maxwell}) are equivalent to Eq.~(\ref{potential0}). 
Provided an initial equilibrium configuration (\ref{potential0}) and proper boundary conditions, we numerically integrate these five equations (\ref{pt}-\ref{maxwell}) to obtain five variables $n_{e}(t,r),\epsilon_{e}(t,r),P_{e}(t,r), \bar p(t,r)$ and $E(t,r)$ describing the electric processes 
around the baryon core boundary. 

\comment{ 
case $n_p\sim\theta(R_c-r)$ (where $R_c$ is the core radius) and
untra-relativistic approximation, Eq.~(\ref{potential0}) has
analytical solutions for the electron number-density $n^0_e$ and
electric potential $V(r)$, and electric fields are over critical
$(E>E_c=m^2_ec^3/(e\hbar))$.
}

\comment{
\begin{figure}
\begin{center}
\includegraphics[width=2.5in]{ev.eps}
\includegraphics[width=2.4in]{nep.eps}
\caption{Left panel: electric potential around the surface of
compact stars; Right panel: electron (red line) and proton (blue
line) distributions. The $\xi$ here we use is a dimensionless
distance from the surface of cores:
$\xi=(12/\pi)^{1/6}\frac{\sqrt{\alpha}}{\Delta}(r-R_c)/\lambda_\pi$, where $\lambda_\pi=(\hbar/m_\pi
c)$ and $\Delta^{-3}=\lambda_\pi^3 n_p(4\pi/3).$} \label{ev}
\end{center}
\end{figure}
}

\section{\bf Oscillations of electron fluid and electric field.}\label{s-osci}

\begin{figure}
\begin{center}
\includegraphics[width=2.5in]{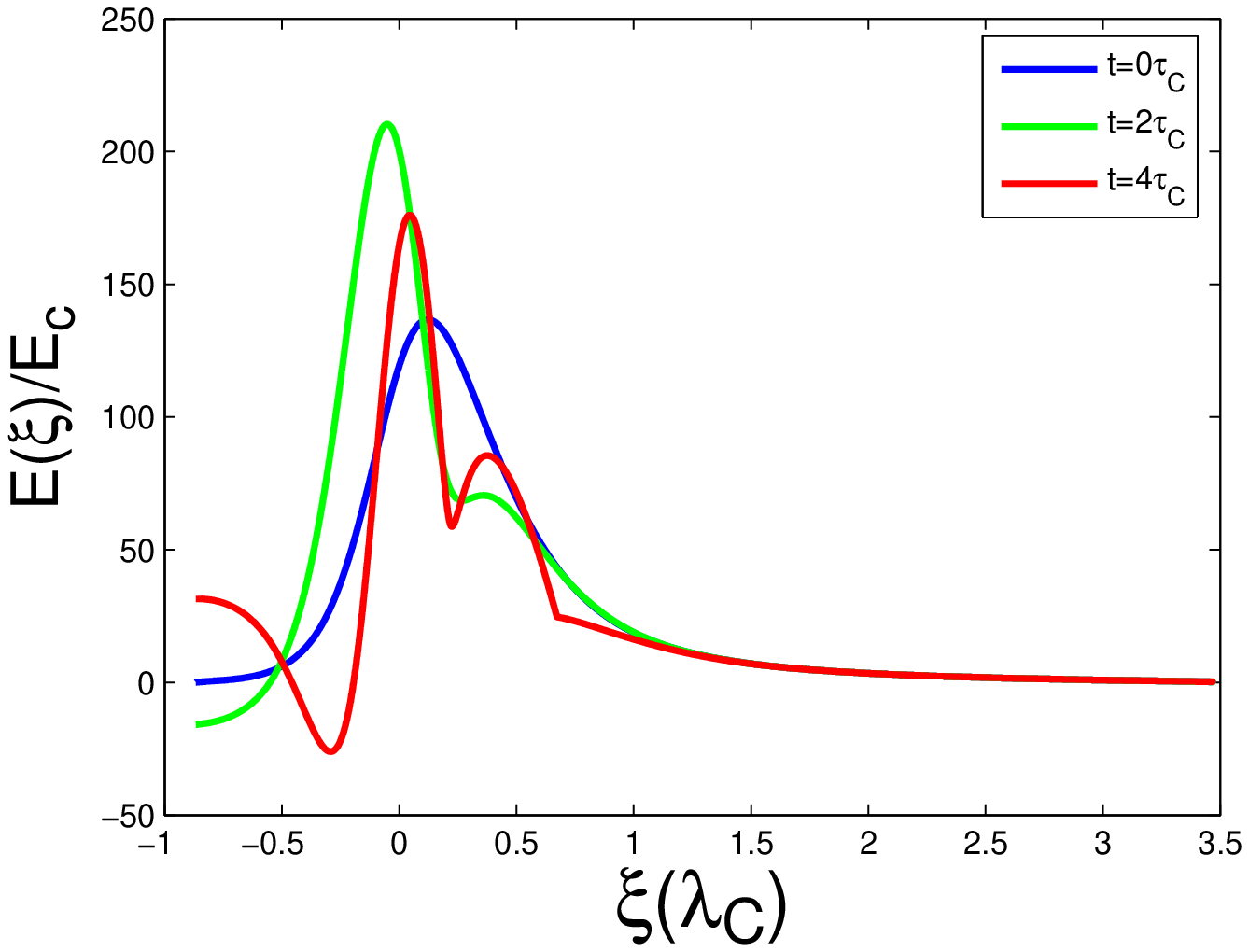}
\includegraphics[width=2.5in]{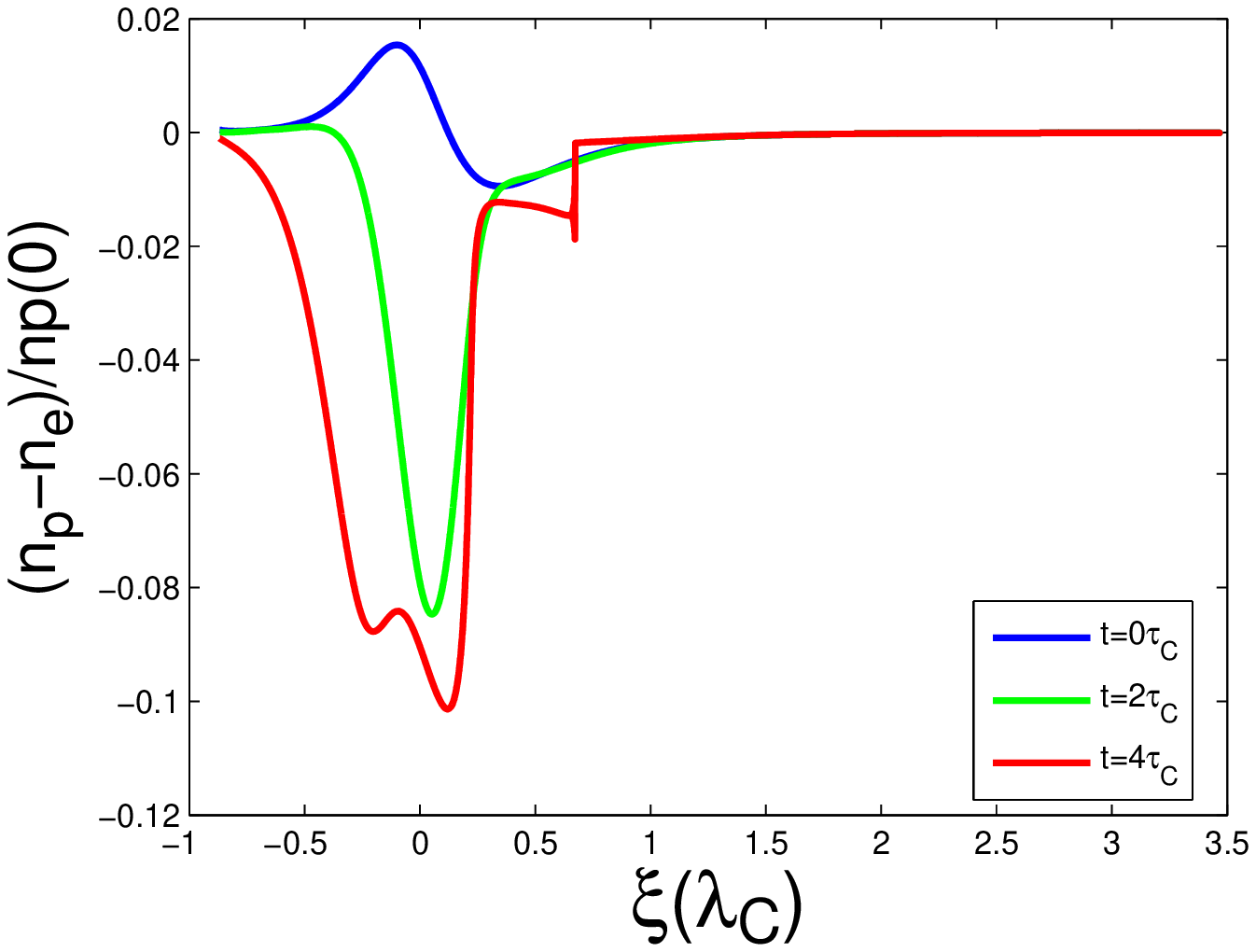}
\caption{The space and time evolution of the electric field (left) and charge-separation (right) around the boundary layer  of the baryon core, $M=10 M_\odot$, 
$R_c \approx 10^7$cm, 
and  $v_p=0.2c$. The coordinate is $\xi\equiv r-R_c$.}
\label{E3t}
\end{center}
\end{figure}

\begin{figure}
\begin{center}
\includegraphics[width=2.3in]{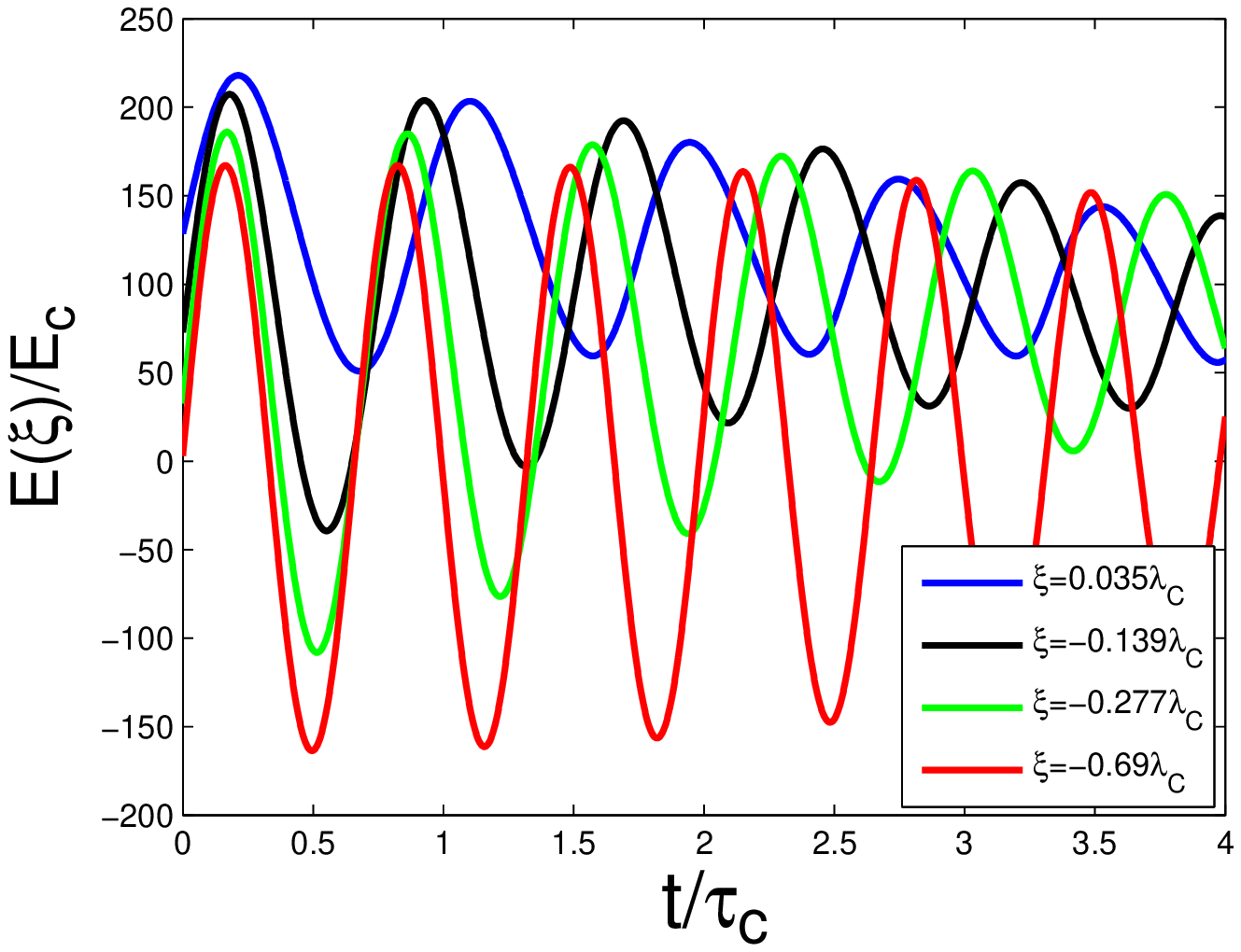}
\caption{Time evolution of electric fields at different radial positions around the boundary layer of the baryon core, $M=10 M_\odot$, 
$R_c\approx 10^7$cm 
and  $v_p=0.2c$. The coordinate is $\xi\equiv r-R_c$.} \label{Etvt}
\end{center}
\end{figure}

We consider the baryon core of mass $M= 10 M_\odot$ and radius 
$R_c \sim 10^7$cm 
at the nuclear density $n_{\rm nucl}$, and select its boundary velocity $v_p=0.2 c$ to represent possible dynamical perturbations of baryon cores. In the proper frame of a rest observer at the core radius $R_c$,
where $g_{tt}(R_c)\approx g^{-1}_{rr}(R_c)$,
we chose the surface
layer boundaries $\xi_-\approx-\lambda_e,
\xi_+\approx3.5\lambda_e$, at which $E_{\rm eq}(\xi_\pm)\approx 0$ and
proper thickness $\ell=\xi_+-\xi_-$, and numerically integrate
Eqs.~(\ref{pt}-\ref{maxwell}) for the electron fluid. Numerical results are presented in Figs.~\ref{E3t} and \ref{Etvt}, showing
that total electric field 
\begin{eqnarray}
E(t,r)=E_{\rm eq}(r)+\tilde E(t,r),
\label{e-field}
\end{eqnarray}
where electron number density, 
energy density, and pressure oscillate around their equilibrium configurations \cite{hrx2010}. 
This is due to the fact that electrons do not possess the strong interaction and their mass is much smaller than the baryon one, as a result, the current density of electron fluid in the boundary layer does not exactly follow the baryon core current density (\ref{cur0}). Instead, triggered by the baryon core current (\ref{cur0}), total electric fields $E(t,r)$ deviate from $E_{\rm eq}(r)$ and increase, which breaks the equilibrium condition (\ref{potential0}), namely, the balance between pressure and  electric force acting on electrons, $d P_e^F/dr + e E_{\rm eq}=0$. Accelerated by increasing electric fields, electrons outside the core start to move inwards following the collapsing baryon core. This leads to the increase of the electron pressure (\ref{pt}) and the decrease of the electric fields. On the contrary, increasing electron pressure pushes electrons backwards, and bounces them back. Overcritical electric fields work against the pressure of ultrarelativistic electrons.  As a consequence, oscillations with frequency $\omega = \tau^{-1}_{\rm osci}\sim 1.5 m_e$ around the equilibrium configuration take place in a thin layer of a few Compton lengths around the boundary of baryon core. These are the main results presented in this article. We would like to point out that these results should not depend on the boundary properties (\ref{baryon}) and (\ref{cur0}) that we assume  for the dynamical perturbations of baryon cores.  The reason is that both electron and proton fluids in baryon cores are at or over nuclear density, and their Fermi momenta are the order of the pion mass $m_\pi$; therefore, electric fields must be at or over critical value $E_c=m_e^2/e$ to do work against motion of charge separation between positively charged baryon and electron fluids, and the frequency of oscillation because of the backreaction should also be the order of $m_e$. It is worthwhile  that these results are further checked by full numerical calculations without assuming the boundary properties (\ref{baryon}) and (\ref{cur0}) of baryon cores, which undergo the dynamical perturbations caused by the gravitational collapse or pulsation.   

\comment{ 
this is quite a contrast to oscillation frequency increasing as electric fields that  discussed in the pressure-less gas of electron-positron pairs in spatial inhomogeneous electric fields \cite{hrx2010}.
}
 
Suppose that the dynamical perturbation of the baryon core is caused by either the gravitational collapse or pulsation of the baryon core, that gains the gravitational energy. Then, in this oscillating process, energy transforms from the dynamical perturbation of the baryon core to the electron fluid via an oscillating electric field.
This can been seen from the energy conservation (\ref{coeqns1}) along a flow line of the electron fluid for $v_e\not= v_p$
\begin{eqnarray}
U^\mu_e(T^{\nu}_{\,\,\,\mu})_{;\nu} =e\bar n_p F_{\mu\nu}U^\mu_eU^{\nu}_B =e\bar n_p\gamma_e\gamma_B(v_p-v_e)g_{rr}E,
\label{decay}
\end{eqnarray}
although we have not yet explicitly proved it.
The energy density of the oscillating electric field is 
\begin{eqnarray}
\epsilon_{\rm osci}\equiv [E^2(t,r)- E^2_{\rm eq}(r)]/(8\pi).
\label{osci_e}
\end{eqnarray}
The energy densities of the oscillating electric field and electron fluid are converted from one to another in the oscillating process with frequencies $\omega \sim \tau^{-1}_{\rm osci}\sim 1.5 m_e$ around the equilibrium configuration. However, the oscillating electron fluid has to relax to the new equilibrium configuration determined by Eqs.~(\ref{potential0}) and (\ref{b_con}) with a smaller baryon core radius $R_c'<R_c$. As a result, the
oscillating electric field must damp out and its lifetime $\tau_{\rm relax}$ is actually a relaxation time to the new equilibrium configuration. As shown in Fig.~\ref{Etvt} the relaxation rate $\tau^{-1}_{\rm relax}\sim 0.05 m_e$. 
We notice very different time scales of strong interacting processes, electric interacting processes and dynamical perturbations of baryon cores: $\tau^{-1}_{\rm stro}\gg \tau^{-1}_{\rm osci}\gg\tau^{-1}_{\rm relax}\gg \tau^{-1}_{\rm coll}$.
\comment{
}

Moreover, when $E(r,t)>E_{\rm eq}(r)$ (see Fig.~\ref{E3t}), the unoccupied
electronic energy-level can be obtained by \cite{Hagen} 
\begin{eqnarray}
{\mathcal E}_{\rm  unocuppied} &=& e\int  g_{rr}^{1/2} dr E(t,r) - {\mathcal E}_{\rm ocuppied}\no\\
&=&e\int g_{rr}^{1/2}dr \tilde E(t,r),
\label{unocu}
\end{eqnarray}
see Eq.~(\ref{occupy}).
This leads to pair production in strong electric fields and converts electric energy into the energy of electron-positron pairs, provided the pair-production rate $\tau_{\rm pair}^{-1}$ is faster than the oscillating frequency $\omega=\tau^{-1}_{\rm osci}$. Otherwise, the energy of oscillating electric fields would completely be converted into the electrostatic Coulomb energy of the new equilibrium configuration of electron fluid, which cannot be not radiative. 

\comment{While, in the phase $E(r)< E_{\rm eq}(r)$, i.e., $\xi_Q < \xi_Q^{\rm eq}$,
when electrons are more compressed, there are no phase-space available for pair-productions, electric energy goes to increase of electron energy.
}

\section{\bf Electron-positron pair production}\label{s-pair}

We turn to the pair-production rate 
in spatially inhomogeneous and temporally oscillating electric fields $E(t,r)$. Although the oscillating frequency $\omega$ is rather large,  the pair-production rate $\tau^{-1}_{\rm pair}$ can be even larger due to the very strong electric fields $E(t,r)$. The pair-production rate can be approximately calculated by the formula for static fields.
The validity of this approximation is justified (see \cite{phreport,Brezin}) by the adiabaticity
parameter $\eta^{-1}=(\omega/m_e)(E_{c}/E_{\rm max})\ll 1$, where $E_{\rm max}$ is the maximal value of the electric field on the baryon core surface $r\simeq R_c$.
Therefore we adopt Eqs.~(38) and (39) and (64)-(66) in Ref.~\cite{Hagen} for the Sauter electric field to estimate the density of the pair-production rate in the proper frame at the core radius $R_c$
\begin{eqnarray}
{\mathcal R}_{\rm pair}
\approx  \frac{e^2 E \tilde E}{4\pi^3
\,\bar G_0(\sigma)
}
e^{-\pi ( E_c/{E})G_0(\sigma)}\sim \frac{e^2 E \tilde E}{4\pi^3
},
\label{3drate}
\end{eqnarray}
where $\tilde E$ (instead of $E$) in the prefactor accounts for the unoccupied electric energy levels, $G_0(\sigma)\rightarrow 0$ and
$\bar G_0(\sigma)\rightarrow 1$ for $\sigma =(\ell /\lambda_e)(E/E_c)\gg 1$. The electron-positron
pairs screen the oscillating field $\tilde E$ so that the number of pairs can be estimated by ${\mathcal N}_{\rm pair}\approx 4\pi R_c^2(\tilde E/e)$. The pair-production rate is $\tau_{\rm pair}^{-1}\approx {\mathcal R}_{\rm pair}(4\pi R_c^2\ell)/{\mathcal N}_{\rm pair}\sim\alpha m_e(\ell/\lambda_e) (E/E_c)\simeq 6.6 m_e > \tau^{-1}_{\rm osci}$. 
The number density of pairs is estimated by $n_{\rm pair}\approx {\mathcal N}_{\rm pair}/(4\pi R_c^2\ell)$. Assuming the energy density $\epsilon_{\rm osci}$ of oscillating fields is totally converted into the pair energy density, we have the pair mean energy $\bar \epsilon_{\rm pair} \equiv \epsilon_{\rm osci}/n_{\rm pair}$. 
Using the parameters $v_p\approx 0.2c$, $R_c\approx 10^{7}$cm, and $M=10M_\odot$, we obtain $\epsilon_{\rm osci}\approx 4.3\times 10^{28}\,{\rm ergs}/{\rm cm}^3, n_{\rm pair}\approx 1.1\times 10^{33}/{\rm cm}^3$, and $\bar \epsilon_{\rm pair} \approx 24.5$MeV. 
These estimates are preliminary without considering the efficiency of pair-productions, 
possible suppression due to strong magnetic fields, and possible
enhancement due to finite temperature effect.  

\section{Gravitational collapse and Dyadosphere}

Up to now, we have not discussed how the dynamical perturbations of baryon cores can be caused by either the gravitational collapse or pulsation of baryon cores. Actually, we have not been able to completely integrate the dynamical equations discussed in Sec.~\ref{basic} for the reasons discussed in Secs.~\ref{int_s} and \ref{model}. Nevertheless, we attempt to use the results of electric field oscillation and pair production obtained in Secs.~\ref{s-field}, \ref{s-osci} and \ref{s-pair} to gain some physical insight into what and how electric processes could possibly occur in the gravitational collapse of baryon cores. 
For this purpose and in order to do some quantitative calculations, we first model the gravitational collapse of baryon cores by the following assumptions:
\begin{enumerate}

\item 
the gravitationally collapsing process is made of the sequence of events (in time) occurring at different radii $R_c$ of the baryon core;

\item 
at each event the baryon core maintains its density profile and sharp boundary as described by Eqs.~(\ref{baryon}) and (\ref{cur0}).
\end{enumerate}
The first assumption is based on the arguments that (i) in the electric processes discussed in Sec.~\ref{s-osci}, the charge-mass ratio $Q/M$ of the baryon core can possibly be approaching to $1$, then the collapse process of the baryon core is slowing down and its kinetic energy is vanishing because the attractive gravitational energy gained is mostly converted into the repulsive Coulomb energy of the baryon core; (ii) then this Coulomb energy can be possibly converted into the radiative energy of electron-positron pairs as discussed in Sec.~\ref{s-pair}, and the baryon core restarts acceleration by gaining gravitational energy. We have already discussed the second assumption in Secs.~\ref{model} and \ref{s-osci}. Here we want to emphasize that (i) the sharp boundary properties (\ref{baryon}) and (\ref{cur0}) in the second assumption are technically used in order to numerically calculate the dynamics of electron fluid in the thin shell around the baryon boundary (Secs.~\ref{s-field}, \ref{s-osci} and \ref{s-pair}); (ii) in the gravitational collapse or pulsation of neutral stellar cores at or over nuclear density, these sharp boundary properties (\ref{baryon}) and (\ref{cur0}) should be abandoned in a more realistic model of simultaneously integrating dynamical equations of electron and baryon fluids over the entire stellar core at macroscopic scales. This turns out to be much more complicated and we will focus on this study in the future.        

\begin{figure}
\begin{center}
\includegraphics[width=2.3in]{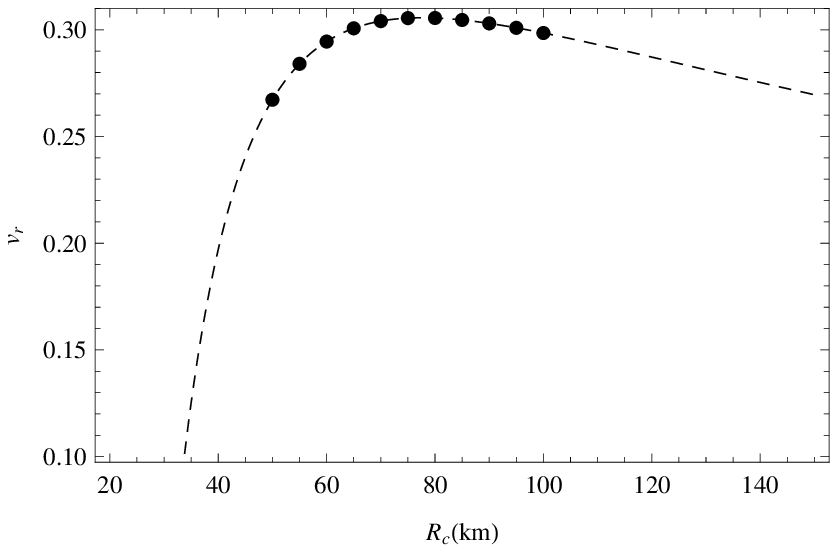}
\caption{The estimate of the core collapsing velocity $v_p\equiv \dot R_c=dR_c/dt$ at different collapsing radii $R_c$ for the baryon core of mass $M=10\, M_\odot$.} \label{c_velocity}
\end{center}
\end{figure}

On the basis of these assumptions, the boundary velocity $v_p(R_c)$ (\ref{v_bar}) and boundary radius $R_c$ [or boundary density $\bar n_{B,p}(R_c)$ (\ref{baryon})] of the baryon core at or over the 
nuclear density are no longer independent parameters, instead they should be related by the gravitational collapse equation of the baryon core. We adopt a simplified model for the gravitational collapse of the baryon core by approximately using the  
collapsing equation for a thin shell \cite{I66,crv2002}
\begin{eqnarray}
\left(\frac{\Omega}{F}\right)^2\left(\frac{dR_c}{dt}\right)^2=\left[1+\frac{GM}{2R_c}(1-\xi_Q^2)\right]^2-1,
\label{coll}
\end{eqnarray}
where at different radii $R_c$ of the baryon core, we define the charge-mass ratio 
\begin{eqnarray}
\xi_Q &\equiv& Q^{\rm eq}/(G^{1/2}M)< 1;\quad Q^{\rm eq} = R^2_c E_{\rm eq},
\label{q/m}
\end{eqnarray}
and 
\begin{eqnarray}
\Omega &\equiv& 1-(M/2R_c)(1+\xi_Q^2)\no\\
F&\equiv & 1-(2M/R_c)+(Q^{\rm eq}/R_c)^2.
\label{o/f}
\end{eqnarray}
The collapsing Eq.~(\ref{coll}) for the collapsing velocity $\dot R_c$ is based on the condition that at each collapsing radius $R_c$, the shell starts to collapse from rest. 
As a result, using these Eqs.~(\ref{coll}-\ref{o/f}) we describe the sequence of events in the gravitationally collapsing process in terms of the collapsing velocities $v_p=\dot R_c=dR_c/dt$ defined by  (\ref{v_bar}) and (\ref{gama_bar}) at different collapsing radii $R_c$ of the baryon core, as shown in Fig.~\ref{c_velocity}.  Thus, at each event the induced inward charged baryon current-density (\ref{cur0}) is given by
\begin{eqnarray}
J_B^r=e\bar n_pU^r_B\approx e\bar n_p(\dot R_c\Omega/F), 
\label{cur}
\end{eqnarray}
as a function of the collapsing radius $R_c$.
The strength of this charged baryon current density (\ref{cur}) depends also on the ratio of the charged baryon number and total baryon number ($N_p/N_B$), which varies in the gravitational collapsing process  because of the $\beta$ processes \cite{mrx2012}. In this article, the $\beta$ processes are not considered and the charged baryon (proton) number $N_p$ is constant; we select two values $N_p/N_B\approx 1/38$ or $N_p/N_B\approx 1/380$ for the charged baryon current density Eq.~(\ref{cur}).
\comment{ 
} 
The collapsing process rate is $\tau^{-1}_{\rm coll}=\dot R_c/R_c \lesssim c/R_c$. If the dynamical perturbation of the baryon core is caused by the gravitational core pulsation, the pulsation frequency can be expressed as $\omega_{\rm pulsa}\simeq \dot R_c/R_c=\tau_{\rm coll}$.

\begin{figure}
\begin{center}
\includegraphics[width=2.5in]{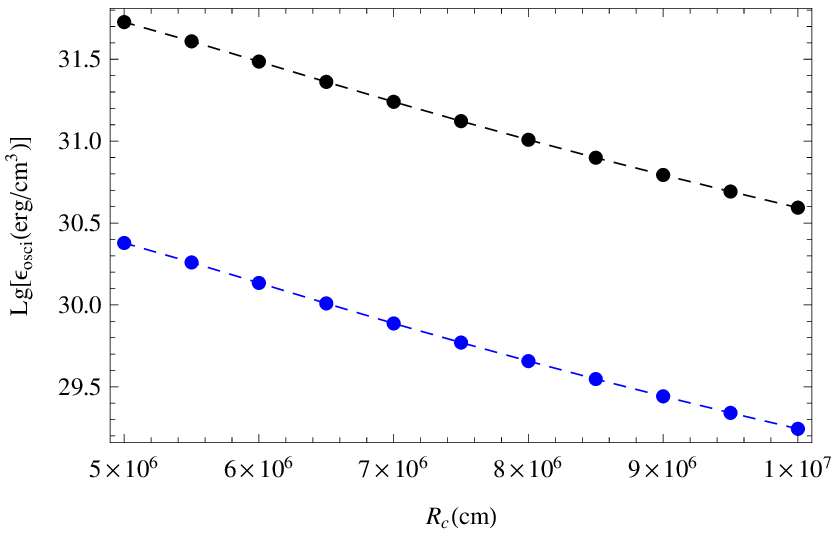}
\includegraphics[width=2.5in]{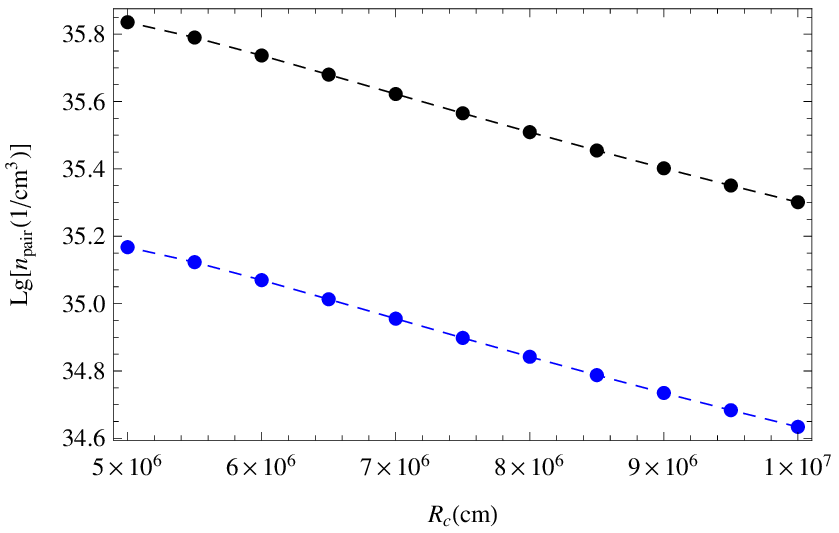}
\caption{The energy (left) and number (right) densities of electron-positron pairs at selected values of collapsing radii $R_c$ for $M=10M_\odot$ and $N_p/N_B\approx 1/38\, ({\rm upper}); 1/380\, ({\rm lower})$. We select $R^{\rm max}_c\sim 10^7{\rm cm}$ so that $\bar n_B\sim n_{\rm nucl}$. }
\label{density}
\end{center}
\end{figure}

In the sequence of the gravitationally collapsing process, at each event characterized by $[R_c, v_p(R_c)]$, we first solve Eqs.~(\ref{potential0}) and (\ref{b_con}) of the equilibrium configuration to obtain the number density ($n_e^{\rm eq}$) and electric field ($E_{\rm eq}$) as the initial configuration of the electron fluid and electric field. Then, with this initial configuration we numerically solve the dynamical equations  (\ref{pt}-\ref{maxwell}) to obtain the dynamical evolution of electron fluid and electric field within the thin shell (a few Compton lengths) around the baryon core boundary, described by Eqs.~(\ref{baryon}) and (\ref{cur}). As a result, based on the analysis presented in Sec.~\ref{s-pair} we calculate the energy and number densities of the electron-positron pairs produced at each event in the sequence of the gravitationally collapsing process. These results are
plotted in Figs.~\ref{density}. Limited by numerical methods, we cannot do calculations for smaller radii. 

\begin{figure}
\begin{center}
\includegraphics[width=2.3in]{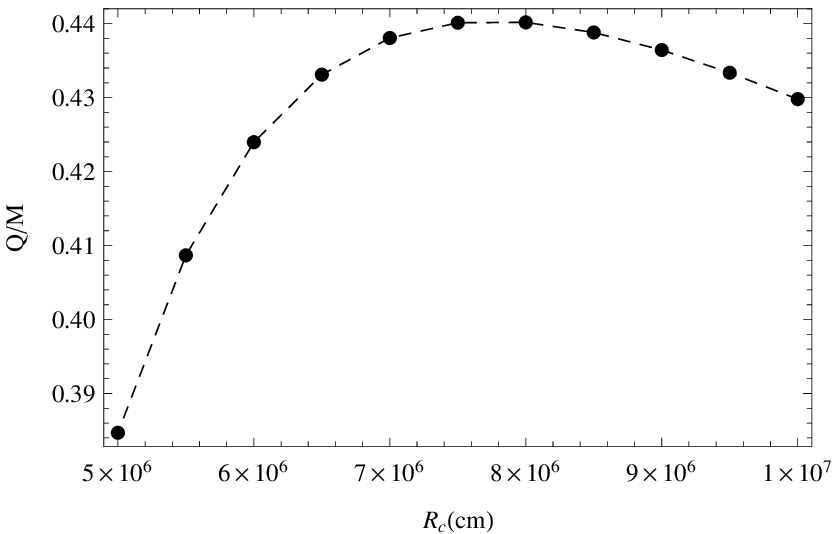}
\caption{The charge-mass ratio $Q/M$ averaged over oscillations of electric fields is plotted at different collapsing radii $R_c$ for the baryon core of mass $M=10\, M_\odot$.} \label{Q/M}
\end{center}
\end{figure}

In addition, at each event in the sequence of the gravitationally collapsing process,
using the Gauss law, $Q=R_c^2 E$, we calculate the charge-mass ratio $Q/M$ averaged over oscillations of electric fields, $Q/M< 1$ as shown in Fig.~\ref{Q/M}. The averaged charge-mass ratio $Q/M$ is not very small, rather about $0.4$ (see Fig.~\ref{Q/M}), implying the possible validity of the first assumption we made that the gravitational collapsing process is approximately made of a sequence of events. In principle, at $Q/M=1$ the gravitational collapsing process should stop, whereas the gravitational collapsing process is continuous for $Q/M=0$ without considering electric interactions. 

It is clear that the ratio $N_p/N_B$ becomes larger, the charged baryon current density (\ref{cur0}) or (\ref{cur}) becomes larger, and all effects of electrical processes we discussed in Secs.~\ref{s-field}, \ref{s-osci} and \ref{s-pair} become larger. 
As shown in Figs.~\ref{density}, for the ratio $N_p/N_B\approx 1/38$, the energy density of electron-positron pairs is about $10^{31}\,{\rm ergs}/{\rm cm}^3$, and the number density of electron-positron pairs is about $10^{35.6}\,/{\rm cm}^3$. The mean energy of electron-positron pairs is $\bar \epsilon_{\rm pair} \equiv \epsilon_{\rm osci}/n_{\rm pair}\sim 10$--$50$ MeV. While, for the ratio $N_p/N_B\approx 1/380$, the energy density of electron-positron pairs is about $10^{30}\,{\rm ergs}/{\rm cm}^3$, the number density of electron-positron pairs is about $10^{34.6}\,/{\rm cm}^3$, and the mean energy of electron-positron pairs $\bar \epsilon_{\rm pair} \equiv \epsilon_{\rm osci}/n_{\rm pair}\sim 10$--$50$ MeV does not change very much.
\comment{  
}

It this article, it is an assumption that the gravitationally collapsing process is represented by the sequence of events: the baryon core starts to collapse from rest by gaining gravitational energy, the increasing Coulomb energy results in decreasing kinetic energy and slowing down the collapse process, the electric processes discussed in Secs.~\ref{s-field}, \ref{s-osci} and \ref{s-pair} convert the Coulomb energy into the radiative energy of electron-positron pairs, and as a result the baryon core restarts to accelerate the collapse process by further gaining gravitational energy. This indicates that in the gravitationally collapsing process, the gravitational energy must be partly converted into the radiative energy of electron-positron pairs. However, we have not been able so far to calculate all processes with very different time and length scales from one event to another in the sequence, so that it is impossible to quantitatively obtain the rate of the conversion of the gravitational energy to the energy of electron-positron pairs.  
Nevertheless,     
by summing over all events in the sequence of the gravitationally collapsing process, we approximately estimate 
the total number and energy of electron-positron pairs produced in the range $R_c\sim 5\times 10^5 -10^7$cm: $10^{56}$--$10^{57}$ and 
$10^{52}$--$10^{53}$ erg for the ratio $N_p/N_B\approx 1/38$; $ 10^{55}$--$10^{56}$ and $ 10^{51}$--$10^{52}$ erg for the ratio $N_p/N_B\approx 1/380$. 
These electron-positron pairs undergo the plasma oscillation in strong electric fields and annihilate to photons to form a neutral plasma of photons and electron-positron pairs \cite{Luca}. This is reminiscent of a sphere of electron-positron pairs and photons, called a Dyadosphere that is supposed to be dynamically created during gravitational collapse in Refs.~\cite{dyado1}.

\section
{\bf Summary and remarks.}
In the simplified model for the baryon cores of neutral compact stars, we show possible electric processes
for the production of electron-positron pairs within the thin shell (a few Compton lengths) around
the boundary of baryon cores that undergo gravitationally collapsing or pulsating processes, depending on the balance between attractive gravitational energy and repulsive electric and internal energies (see the numerical results in Ref.~\cite{G_2005}). This indicates a possible mechanism that the gravitational energy is converted into the energy of electron-positron pairs in either baryon core collapse or pulsation. 

In theory, this is a well-defined problem based on the Einstein-Maxwell equations, particle-number and energy-momentum conservation (\ref{coeqns1})-(\ref{etensor}), and equations of states, as well as the Sauter-Heisenberg-Euler-Schwinger mechanism. However, in practice, it is a rather complicated problem that one has to deal with various interacting processes with very different time and length scales. The approach we adopt in this article is the adiabatic approximation: the interacting processes with very small rates are considered to be adiabatic processes in comparison with the interacting processes with very large rates \cite{adiab-note}. Therefore, we try to split the problem of rapid microscopic processes from the problem of slow macroscopic processes, and focus on studying rapid microscopic processes in the background of adiabatic (slowly varying) macroscopic processes.  
The adiabatic approximation we adopted here is self-consistently and quantitatively justified by process rates 
\begin{eqnarray}
\tau^{-1}_{\rm strong}\gg \tau^{-1}_{\rm pair} > \tau^{-1}_{\rm osci} \gg \tau^{-1}_{\rm relax}\gg \tau^{-1}_{\rm coll}, 
\label{adia}
\end{eqnarray}
studied in this article.
In addition to the adiabatic approximation, we have not considered in this over simplified model the hydrodynamical evolution of baryon cores, the back-reaction of oscillations and pair-production on the collapsing or pulsating processes, and the dynamical evolution of the electron-positron pairs and photons. Needless to say, these results should be further checked by numerical algorithms
integrating the full Einstein-Maxwell equations and proper EOS of particles in gravitational collapse. Nevertheless,
the possible consequences of these electromagnetic processes discussed in this article are definitely interesting and could be possibly relevant and important for understanding  energetic sources of supernovae and gamma-ray bursts.

\section{\bf Acknowledgment}

R.~Ruffini and S.-S.~Xue are grateful to T.~Damour, G.~t'Hooft, H.~Kleinert, G.~Preparata and other colleagues for many discussions on this issue in the last 14 years. Authors thank R.~Jantzen for reading the manuscript. W.-B. Han is now supported
by the Chinese NSFC (No.11243002).

\end{document}